\documentclass[aps,prl,twocolumn,preprintnumbers]{revtex4}

\usepackage{psfrag,graphicx}
\usepackage{dcolumn}
\usepackage{amsmath,amssymb}
\usepackage{bm}
\usepackage{amsfonts,amssymb,amsmath}        

\newcommand{\be}{\begin{equation}}
\newcommand{\ee}{\end{equation}}
\newcommand{\bq}{\begin{eqnarray}}
\newcommand{\eq}{\end{eqnarray}}
\newcommand{\Sp}{\,\,\,\,\,\,}

\newcommand{\ket}[1]{\left | \, #1 \right\rangle}

\begin{document}

\title{Quantum computation with abelian anyons on the honeycomb lattice}
\date{\today}

\author{Jiannis K. \surname{Pachos}}
\affiliation{DAMTP, University of Cambridge, Wilberforce Road, Cambridge
CB3 0WA, UK}

\begin{abstract}

We consider a two-dimensional spin system that exhibits abelian anyonic
excitations. Manipulations of these excitations enable the construction of a
quantum computational model. While the one-qubit gates are performed
dynamically the model offers the advantage of having a two-qubit gate that
is of topological nature. The transport and braiding of anyons on the
lattice can be performed adiabatically enjoying the robust characteristics
of geometrical evolutions. The same control procedures can be used when
dealing with non-abelian anyons. A possible implementation of the
manipulations with optical lattices is developed.

\end{abstract}

\maketitle

Topological models address in the most sufficient way the problem of
decoherence, which is the main obstacle in the realization of quantum
computation~\cite{Kitaev_97,Freedman,Mochon,Bonesteel}. This is
achieved by encoding information in
the statistical properties of non-abelian anyons. Hence, information remains
unaffected by temperature or local perturbation as long as the nature of the
anyons does not change. Subsequently, information can be processed by
braiding anyons, an evolution which is topologically protected from control
errors. Nevertheless, due to their exotic nature these models are hard to
realize physically. To approach this problem we construct a simple model for
quantum computation with topological characteristics that can be implemented
easier than the fully topological quantum computer. The simplest case is to
identify the topological phase arising from the statistics of abelian anyons
with a two-qubit gate~\cite{Lloyd}. In this scenario qubit encoding and
one-qubit gates can be performed in the usual dynamical way.

The purpose of this letter is twofold. First, we propose a physically
realizable model for the generation and detection of abelian
anyons. The latter can be subsequently used to perform quantum
computation that has a topologically protected
two-quit gate. Second, we develop the test ground for the generalization to
the more demanding non-abelian case where topological quantum memory or
computation can be performed. Indeed, the main ingredients of the abelian
case, such as the creation of anyons, their braiding and their subsequent
measurement can be carried through to the case of non-abelian anyons.

\begin{center}
\begin{figure}[ht]
\resizebox{!}{3.3 cm}
{\includegraphics{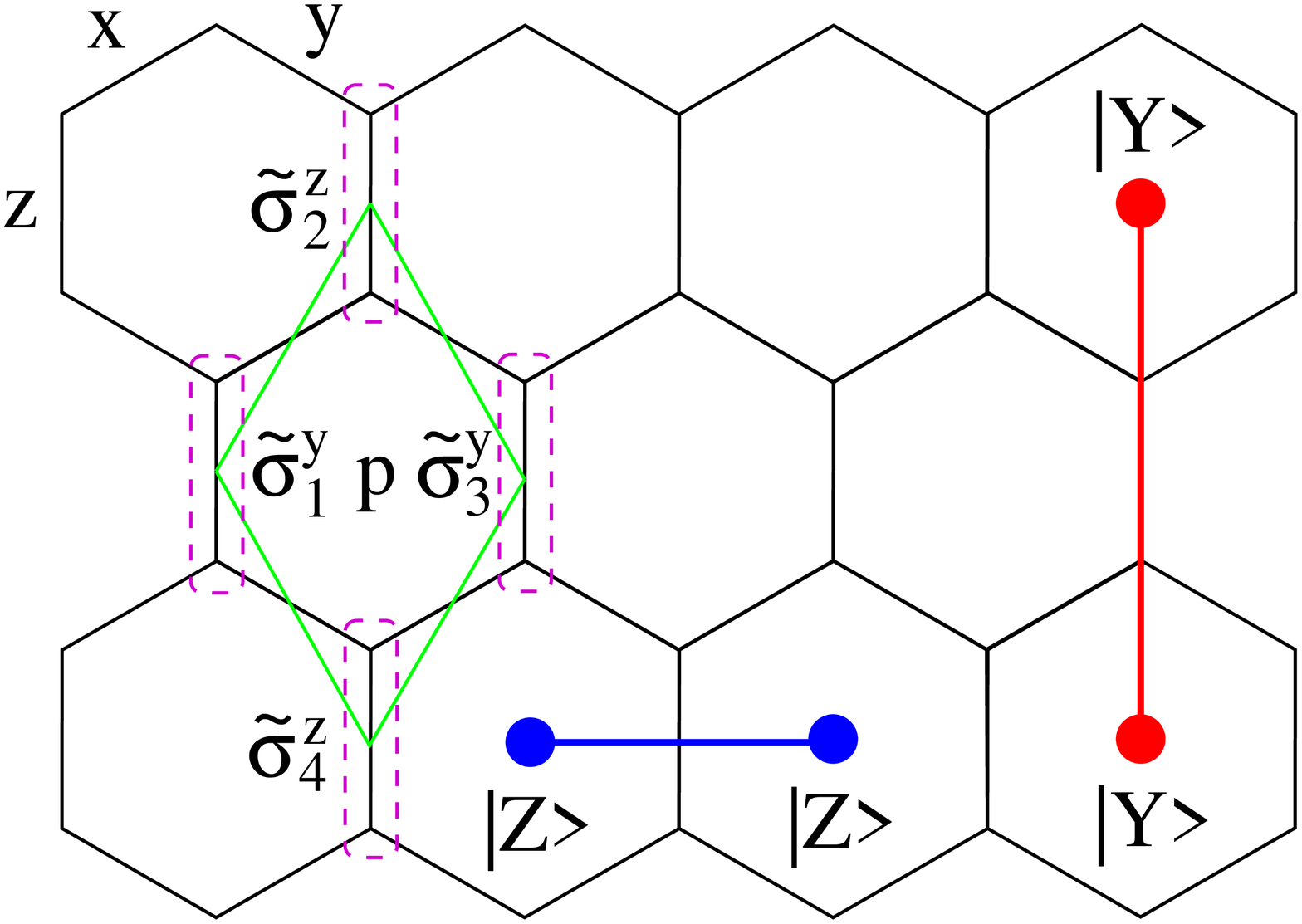}}
\caption{\label{romve} The honeycomb lattice with the three different
types of links $x$, $y$ and $z$. The distribution of the four spin
interaction, $Q_p$, is indicated. Possible positions of the excitation pairs
produced by rotations of effective spins.}
\end{figure}
\end{center}

While the ideas here can be applied to a variety of different
setups~\cite{Paredes,Bombin} we shall employ a particular physical model,
namely a two dimensional spin-1/2 system where the spins are located at the
vertices of a honeycomb lattice~\cite{Kitaev}. The spins are set to interact
with each other via the following Hamiltonian
\be
H = -J_x \sum_{\text{x-links}} \sigma_j^x \sigma_k^x
-J_y\sum_{\text{y-links}} \sigma_j^y \sigma_k^y -J_z\sum_{\text{z-links}}
\sigma_j^z \sigma_k^z
\label{Ham}
\ee
where ``x-links", ``y-links" and ``z-links" label the three different link
directions of the honeycomb lattice as depicted in Figure~\ref{romve}. It
has been shown~\cite{Kitaev} that this model exhibits two main distinctive
phases. For $|J_x|+|J_y|\ge |J_z|$, $|J_y|+|J_z|\ge |J_x|$ and
$|J_z|+|J_x|\ge |J_y|$ a gapless sector appears that supports non-abelian
excitations, while the violation of any of the three inequalities leads to a
gap. Here we shall focus on the latter phase for a concrete choice of
couplings, namely $|J_z|>|J_x|+|J_y|$. As we shall explicitely see here this
phase is equivalent to Kitaev's toric code which exhibits three
distinguished species of particle excitations above the ground state $0$,
namely a fermion, $\epsilon$, and two bosons, the chargeon, $e$, and the
magnon, $m$~\cite{Kitaev_97}. Interestingly, exchanges between different
particles reveals their anyonic character.

To identify the properties of the abelian phase in the honeycomb lattice it
is instructive to employ perturbation theory~\cite{Kitaev}. Nevertheless,
the discrete qualitative properties of the model will stay the same even
beyond the perturbative approximation~\cite{Jiannis}. Consider the
coupling regime where 
$J_x,J_y,J_z\ge 0$ and $J_z\gg J_x,J_y$. The dominant $z$-interaction has as
lower eigenstates $\ket{\tilde\uparrow} \equiv\ket{\uparrow\uparrow}$ and
$\ket{\tilde\downarrow} \equiv \ket{\downarrow\downarrow}$ for each
$z$-link. The Pauli operators that act on these states are given by
$\tilde\sigma^x\equiv \sigma^x\sigma^x$, $\tilde\sigma^y \equiv
\sigma^y
\sigma^x \approx \sigma^x \sigma^y$ and $\tilde \sigma^z \equiv
\sigma^z 1 \approx 1 \sigma^z$, where `$\approx$' stands for equality
in the space spanned by the $\ket{\tilde \uparrow}$ and $\ket{\tilde
\downarrow}$ vectors. Considering up to fourth order
in perturbation theory, the Hamiltonian (\ref{Ham}) becomes
\be
H \simeq -{J_x^2J_y^2 \over 16J_z^3} \sum_p \tilde \sigma^y_1
\tilde \sigma^z_2 \tilde\sigma^y_3 \tilde\sigma^z_4 \equiv
 -J_{\text{eff}} \sum_p Q_p,
\label{Ham_romve}
\ee
where an irrelevant constant term has been omitted. The summation in
(\ref{Ham_romve}) is over all plaquettes of the lattice, while the
lower indices of the Pauli operators denote the position of the
effective spins around the plaquette, $p$, starting from the left and
moving clockwise (see Figure~\ref{romve}).

This Hamiltonian is unitarily equivalent to the more familiar Kitaev's toric
code~\cite{Kitaev_97}, but here we shall work with Hamiltonian
(\ref{Ham_romve}). Before studying the particle excitations we first
determine the ground state, $\ket{0}$. Since $Q_p^\dagger = Q_p$ and
$Q_p^2=1$, the eigenvalues of $Q_p$ are $\pm 1$ and since $[Q_p,Q_{p'}]=0$
for all $p$, $p'$, the lowest energy state has to satisfy
$Q_p\ket{0}=\ket{0}$ for all $p$. Excitations of Hamiltonian
(\ref{Ham_romve}) are given by relaxing this requirement at some plaquettes.
For periodic boundary conditions we have $\prod_pQ_p=1$ at all times, hence,
excitations should appear in pairs. For open boundary conditions single
excitations are possible, where the other part of the pair can be considered
to be outside the system. Nevertheless, for manipulations away from the
boundary we may assume that the generated particles come in pairs.

\vspace{-0.5cm}
\vspace{-1cm}

Consider a rotation $\tilde\sigma^z$ applied on a certain $z$-link. As
$\tilde\sigma^z$ anticommutes with the $Q_p$ interactions of the
plaquettes on its left and on its right it will increase the energy of
these plaquettes
by $\Delta E = 2J_\text{eff}$. Similarly, if we perform a rotation
$\tilde\sigma^y$, then the energy of the plaquettes above and below
increases by $\Delta E$. If a rotation $i\tilde\sigma^x$ is performed, then
all four neighboring plaquettes obtain an energy increase of $\Delta E$
each. Hence, the low lying energy eigenstates are elementary pairs of
excitations given by
\be
\ket{Z} = \tilde\sigma_j^z \ket{0} \Sp \text{and} \Sp
\ket{Y} = \tilde\sigma_j^y \ket{0}.
\label{electric}
\ee
These excitations live in appropriate neighboring plaquettes as indicated in
Figure~\ref{romve}. Moreover, one can combine such pairs of excitations in a
chain of $\tilde\sigma^z$ and $\tilde \sigma^y$ rotations to create two
excitations at the end points of an arbitrary string $s$. Similarly, one can
define a pair of fermionic excitations, $\ket{X}$, as
\be
\ket{X[s]} = \prod_{j\in s} i\tilde\sigma_j^x \ket{0},
\label{mag_ferm}
\ee
where $s$ is the path along which $\tilde\sigma^x$ rotations are
performed in such a way as interweaving of the constituent $\tilde
\sigma^y$ and $\tilde \sigma^z$ paths is achieved.
The eigenvalues of the energy of the $Y$ and $Z$ excitations have an energy
gap above the ground state given by $\Delta E = 4J_{\text{eff}}$, while for
the $X$ excitation the energy gap is double. Note that, the eigenvalue of
the energy of each excitation is independent of the length of the string,
$s$, as the contributions to the energy come only from its two end-points.

From these definitions one can easily deduce the fusion rules $X\times X =
Y\times Y = Z\times Z = 1$, $X\times Y= Z$, $Y\times Z = X$ and $Z\times X =
Y$, where $1$ denotes the vacuum. Indeed, two successive identical Pauli
rotations ($X$, $Y$ or $Z$), that correspond to superposing two pairs of the
same excitations, will give the identity operation. The generation of an
$X$-particle out of the fusion of $Y$- and $Z$-particles follows
straightforwardly from the property $\tilde\sigma^y\tilde\sigma^z =
i\tilde\sigma^x$. Furthermore, from definitions (\ref{electric}) and
(\ref{mag_ferm}) one can deduce the statistics of these particles.
Exchanging two $X$-particles results in a complete overlap of their strings
at $2(2k+1)$ sites, where $k$ is an inteeger, giving eventually
$(i\tilde\sigma^x)^2=-1$ as an overall
factor that reveals their fermionic character. To probe the anyonic character
we circulate an $X$-particle around a single $Y$- or $Z$-particle. This loop
that corresponds to two consecutive exchanges gives a string operator that
anticommutes with the string of the $Y$- or $Z$-particles, due to the
anticommuting properties of the Pauli operators at the site of their
intersection. Hence, a minus sign will be produced revealing that the $X$
and $Y$ or $Z$ particles behave as anyons with respect to each other with
statistical angle $\theta=\pi/2$. Note, that due to the undetermined
structure of the strings for the $Y$ and $Z$ particles it is not
straightforward to deduce their statistical properties.

In order to use this setup to perform quantum computation we encode
the states of a qubit in the space of excitations of the
system, namely $\ket{0_L} = \ket{X}$ and $\ket{1_L} =\ket{Y}$ (or $\ket{1_L}
=\ket{Z}$ depending on the position of the excitation). To manipulate
these states we place them in
neighboring plaquettes. A one-qubit rotation, e.g. of the $\ket{0_L}$ state,
can then be performed by an appropriate $z$-rotation,
$e^{-i\tilde\sigma^z\theta}\ket{X} = \cos \theta \ket{X} - i\sin \theta
\ket{Y}$. The $\tilde \sigma^z = \sigma^z 1$ rotation is performed at
the effective spin that connects the pair of $X$ particles. A relative phase
between the two states $\ket{0_L}$ and $\ket{1_L}$ can be generated by
changing the relative self-energy of the corresponding excitations. Hence,
a general one-qubit gate can be obtained in the usual dynamical way.

The realisation of a two qubit gate requires braiding of the anyonic
particles. To transport the anyons around the two dimensional lattice
one can create a
trapping potential well with depth $V$ for each anyon. This can be achieved
by reducing the coupling $J_{\text{eff}}$ by $V$ at the corresponding
hexagon. A simple implementation is by increasing the couplings of the two
$z$-links that surround a particular hexagon occupied by the anyon to the
value $J_z'$. Thus, the effective coupling becomes $J_{\text{eff}}' =
J_x^2J_y^2/(4(J_z+ J_z')^2J_z')$. If $J_z'=3J_z$, then the eigenvalue of the
excitation corresponding to that hexagon is reduced twelvefold generating a
strong and efficient trapping well. The employed interaction,
$\sigma^z\sigma^z$, does not change the nature of the particles,
i.e. it does not correspond to a $\tilde\sigma^x$, $\tilde\sigma^y$ or
$\tilde\sigma^z$ rotations. By moving the trapping potential in an adiabatic
fashion it is possible to move each excitation independently. Hence,
braiding can be performed that can actually reveal the anyonic
behaviour of the particles or can be employed to process information.

\begin{center}
\begin{figure}[ht]
\resizebox{!}{3.5 cm}
{\includegraphics{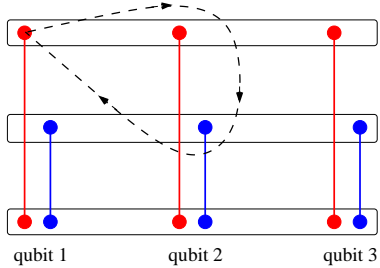}}
\caption{\label{register} A possible arrangement of anyons comprising a
quantum computer. The red particles are the fermions, $X$, while the blue
ones can be either $Y$ or $Z$ particles. The depicted evolution
results in a controlled phase gate.}
\end{figure}
\end{center}

Now we have all the necessary ingredients to build a quantum computer. The
register includes pairs of particles so one can arrange them as in
Figure~\ref{register}. Each qubit is allocated at a specific region on the
lattice where arbitrary superpositions of $X$- and $Y$-particles can be
produced to encode an arbitrary qubit state. To perform a two-qubit gate one
needs to transport the $X$-particle of the control qubit around both, $X$-
and $Y$-particles of the target qubit, as shown in Figure~\ref{register}. As
a result the state $\ket{XY}$ acquires a minus sign while all the other
states remain te same. In the logical space this corresponds to a controlled
phase gate, where a minus sign is generated only for the logical state
$\ket{01}$. This procedure can be performed between any two arbitrary
qubits, which together with the general one-qubit gate result into
universality.
\vspace{-0.5cm}
\vspace{-1.2cm}
The value of the controlled phase gate is fixed by the statistics of the
permuted particles and is resilient to any deformations of the spanned loop
as long as it circulates the target particles once. Moreover, moving the
particles around with the trapping potentials is immune to control errors
that would occur if the transportation was performed by dynamical means. For
the latter case one has to rotate the effective spins successively along the
desired trajectory, with errors that would accumulate at each translation
step. By moving the potential well adiabatically the trapped particle
follows the lowest energy configuration without the need for any dynamical
control of the effective spins and subsequent introduction of control
errors. This resilience which is of geometrical origin gives a great
advantage in transporting and braiding the anyons fault tolerantly.

It has been shown~\cite{Duan_02} that Hamiltonian (\ref{Ham}) can be
generated with a Bose-Einstein condensate of atoms superposed with a two
dimensional optical lattice configuration. In particular, consider two bosonic
species labelled by $\sigma=a, b$, e.g. given by two hyperfine levels of an
atom. They can be trapped by employing two in-phase optical lattices for
each direction. The geometry of the hexagonal lattice can be produced with
three pairs of properly aligned counterpropagating laser
fields~\cite{Duan_02}. The tunnelling of atoms between neighboring sites is
described by $-\sum_{l\sigma} (t_\sigma a_{l\sigma}^\dagger a_{(l+1)\sigma}
+\text{H.c.})$. When two or more atoms are present in the same site, they
experience collisions given by ${1\over 2}\sum_{l\sigma \sigma'}  U_{\sigma
\sigma'} a^\dagger_{l\sigma} a^\dagger_{l\sigma'} a_{l\sigma'} a_{l\sigma}$.
We now consider the limit $t \ll U$ where the system is in the Mott
insulator regime with one atom per lattice site~\cite{Raithel}. In this
regime, we can
employ the pseudospin basis of $|\!\!\uparrow\rangle\equiv
|n_l^a=1,n_l^b=0\rangle$ and $|\!\!\downarrow\rangle\equiv
|n_l^a=0,n_l^b=1\rangle $, for lattice site $l$, and the effective evolution
can be expressed in terms of the corresponding Pauli operators. Consider
initially high enough laser intensities such that each site corresponds to a
free spin $1/2$ particle. By applying Raman transitions between the states
$\ket{a}$ and $\ket{b}$ it is possible to perform an arbitrary local spin
rotation. It is easily verified that when one of the tunnelling coupling is
activated, say $t_a$, the Ising interaction is realized between neighboring
sites~\cite{Pachos1,Kuklov,Duan_02},
\be
H_1=-t_a^2 \Big( \frac{1}{U_{aa}}-\frac{1}{2U_{ab}}\Big) \sum_{<l,m>}
\sigma^z_l\sigma^z_m,
\label{Ham1}
\ee
where $<l,m>$ are the nearest neighbors connected by the
$t_a$ coupling. This procedure can implement the interactions of
the $z$-links of Hamiltonian (\ref{Ham}), where the $J_z$ coupling varies as
a function of $t_a$. To generate the interactions in the $x$- and $y$-links
one needs to create an anisotropy between the $x$ and $y$ spin directions.
For that we activate a tunnelling by means of Raman
couplings~\cite{Pachos1,Duan_02} consisting of two standing lasers $L_1$ and
$L_2$, giving finally the interaction terms $\sum_{<l,m>}
\sigma^x_l\sigma^x_m$ or $\sum_{<l,m>} \sigma^y_l\sigma^y_m$
depending on the phase difference between the laser radiations, $L_1$
and $L_2$. Appropriate combinations of these interactions gives
finally all the necessary terms for realizing Kitaev's Hamiltonian.

Upon this basis we can realize a quantum computational scheme with abelian
anyons. As we have seen, the generation of particle excitations involves the
rotation of effective spins, or equivalently, rotations of the spins at the
$z$-links of the original lattice. In this way initialization of the qubit
space is straightforwardly implemented by local Raman transitions. To
perform a one-qubit rotation an appropriate detuned laser radiation can be
used to induce a local $\sigma^z$ rotation. To perform the trapping of the
excitations we choose to use localized $\sigma^z\sigma^z$ interactions that
would lower the $J_{\text{eff}}$ coupling, thereby reducing the overall
energy of the excitations at a specific plaquette. This can be implemented
simply by focusing a laser field in-between the spins of the $z$-link, that
will reduce the potential barrier thus increasing the $t_a$ tunnelling
(see Eqn. (\ref{Ham1})). Note, that this laser
does not need to have a cross-section smaller than the lattice period,
but merely to have a strong effect on the tunnelling between two
certain sites and weaker between the rest of the sites, thus trapping
the anyonic excitation in that neighborhood. By
moving the focus of this laser around the lattice, the potential minimum and
its trapped particle will also move accordingly. To suppress the generation
of undesired relative phases between the excitations due to fluctuations of
the intensity of the trapping laser one can use the same laser source for
generating the trapping potentials for all the particles. The main source of
decoherence, beyond the leakage of coherence outside the pseudospin space
due to heating of the atoms, is the generation of virtual pairs of particles
due to thermal fluctuations. These excitations can actually move the
particles out of their trapping potentials, which will register as
decoherence of the encoded information. To avoid such problems, we would
like to have the temperature small enough compared to the depth of the
trapping well, $V$, thus exponentially suppressing the generation of virtual
pairs.
\vspace{-0.5cm}
\begin{center}
\begin{figure}[ht]
\resizebox{!}{6.5 cm}
{\includegraphics{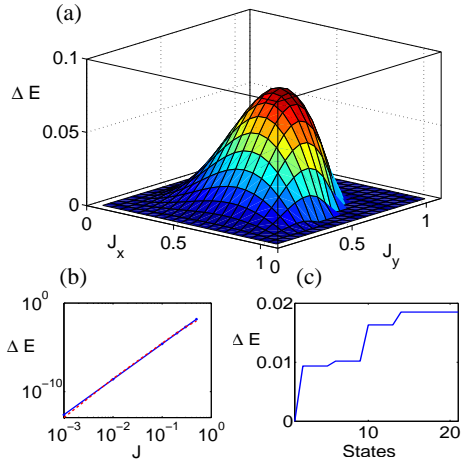}}
\caption{\label{comb2} (a) The numerically obtained energy gap, $\Delta E$,
between the ground and the first excited state with $J_z=1$. (b) Loglog plot
of $\Delta E$ as a function of the coupling $J=J_x=J_y$. There is very good
agreement between the solid (blue) line, predicted by perturbation
theory, and the dashed (red) line, given by numerical
diagonalization. (c) $\Delta E$, for various states above the ground state.
As expected from the theory, $\Delta E$ almost doubles for a certain set of
states that correspond to $X$ excitations compared to $Y$ or $Z$
excitation.}
\end{figure}
\end{center}
\vspace{-1cm}

The measurement procedure required for the presented scheme has to be able
to distinguish between the vacuum state and a particular excitation. As we
have seen, excitations are localized objects that live within the boundaries
of the trapping potentials. A generic measurement should therefore be able
to address the density profile of the particular atomic states as a function
of position and determine their correlations. To increase the
distinguishability between their correlations one may alternatively
use local time modulations of the optical lattice to induce Bragg
scattering resonant between the particular excitation and the first
excited vibrational state of the atoms. By direct population
measurement or by measurement of the atomic correlations one can, in
principle, deduce the state of the
system~\cite{Toth,Altman,Kuklov_99,Grondalski}.

Finally, we would like to see to which extend perturbation theory describes
the initial model by employing numerical diagonalization of Hamiltonian
(\ref{Ham}). Towards that we consider a periodic honeycomb lattice
consisting of 16 spins. Figure~\ref{comb2}(a) presents the energy gap,
$\Delta E$, between the ground and the first excited states as a function of
$J_x$ and $J_y$ where we have set $J_z=1$. It is clear that the model has a
gap for $|J_x|+|J_y|< |J_z|$ as expected from the analytical
treatment~\cite{Kitaev}. Furthermore, it is possible to compare the energy
gap of the perturbative theory with the one obtained numerically for
$J_x,J_y\ll J_z$. Figure~\ref{comb2}(b) presents a comparison between
perturbation (solid, blue line) and the numerical diagonalization (dashed,
red line) of $\Delta E$ for various values of the coupling $J=J_x=J_y$. We
observe that the two curves are in excellent agreement indicating the
validity of perturbation theory for a wide range of coupling values $J$. As
a last test we consider a part of the spectrum above the ground state. We
saw previously that the energy gap of the fermionic excitation, $X$, is
double than the energy gap of $Y$ or $Z$ excitation. Indeed,
Figure~\ref{comb2}(c) gives $\Delta E$ for the first few excitations where
the doubling of $\Delta E$ is clearly observed. Unfortunately, the system of
16 spins that we are able to diagonalize numerically is too small to
represent faithfully the properties of the spectrum expected for a large
system. A complete theoretical study that presents the properties of the
excitation beyond perturbation theory will be presented elsewhere.

{\em Acknowledgements.} We would like to thank Alastair Kay, Seth Lloyd,
John Preskill and J\"org Schmiedmayer for helpful conversations. This work
was supported by the Royal Society.

\end{document}